\def\lesssim{{_ <\atop{^\sim}}}

\def\ap3m{AP$^3$M}
\def\LCDM{$\Lambda$CDM}
\def\hkpc{$h^{-1}{\ }{\rm kpc}$}
\def\hMpc{$h^{-1}{\ }{\rm Mpc}$}
\def\hMsun{$h^{-1}{\ }{\rm M_{\odot}}$}
\def\kms{${\rm{\ }km{\ }s^{-1}}$}
\def\nbody{$N$-body}

\def\ea{et~al.~}                            

\def\lesssim{\mathrel{\hbox{\rlap{\hbox{\lower4pt\hbox{$\sim$}}}\hbox{$<$}}}}
\def\gtrsim{\mathrel{\hbox{\rlap{\hbox{\lower4pt\hbox{$\sim$}}}\hbox{$>$}}}}

\newcommand{\AAA}[3]    {\mbox{A\&A~\textbf{#1},~#2~(#3)}}

\newcommand{\ApJ}[3]    {\mbox{ApJ~\textbf{#1},~#2~(#3)}}

\newcommand{\ApJL}[3]   {\mbox{ApJ~Lett.~\textbf{#1},~#2~(#3)}}

\newcommand{\MNRAS}[3]  {\mbox{MNRAS~\textbf{#1},~#2~(#3)}}
\newcommand{\Nature}[3] {\mbox{Nature~\textbf{#1},~#2~(#3)}}

\newcommand{\Science}[3]{\mbox{Science~\textbf{#1},~#2~(#3)}}

\newcommand{\PhRevD}[3] {\mbox{Phys.~Rev.~\textbf{D#1},~#2~(#3)}}
\newcommand{\astroph}[1]{\mbox{\texttt{astro-ph/#1}}}
\newcommand{\hepph}[1]  {\mbox{\texttt{hep-ph/#1}}}

\documentstyle[epsfig]{mn}

\begin{document}

\title{Bumpy Power Spectra and Galaxy Clusters}

\author[Knebe A., Islam R. \& Silk J.]
       {Alexander Knebe$^{1}$, Ranty R. Islam$^{2}$, and Joseph Silk$^{2}$\\        
       {$^1$Theoretical Physics, Keble Road, Oxford OX1 3NP, UK}\\
       {$^2$Astrophysics, Keble Road, Oxford, OX1 3RH, UK}}

\date{Received ...; accepted ...}

\maketitle

\begin{abstract}
The evolution of the abundance of galaxy clusters is not a reliable
measure of $\Omega$ if there are features on scales of a few Mpc in
the primordial power spectrum. Conversely, if we know the cosmological
model parameters from other measurements, the cluster abundance
evolution permits us to probe features in the power spectrum that are
in the nonlinear regime at the present epoch, and hence difficult to
discern directly from current epoch measurements.  

We have investigated the influence of an artificially introduced
Gaussian feature on an otherwise unperturbed SCDM power spectrum
on scales corresponding to \mbox{$k\sim 0.4-0.8 \ h{\rm Mpc}^{-1}$}.
Using these modified spectra as an input to cosmological
\nbody\ simulations, we are able to show that in terms of the cluster 
abundance evolution, a SCDM model displays characteristics
similar to an OCDM model.  However, strong modifications would also be
visible at a redshift $z=0$ in the dark matter power spectrum whereas
minor alterations to the usual SCDM spectrum are washed away by
non-linear evolution effects.  We show that alterations to the dark
matter power spectrum like those presented in this paper do not leave
any imprint in the present density fluctuation spectrum and the
velocity distribution of galaxy clusters; nearly all models agree with
each other and do not coincide with our fiducial OCDM model,
respectively.  We therefore conclude that features with
characteristics such as discussed here might not be detectable using
observations of the galaxy power spectrum, the local cluster abundance
or the large-scale velocity field as measured by the velocity
distribution of galaxy clusters.

The only quantity that shows a pronounced difference at the present
epoch between our models under investigation is the halo-halo
correlation function which appears to be strongly biased with respect
to an unmodified SCDM model. This is due to a lack of power on certain
scales which subsequently modifies the relative amplitude of high-
and low-$k$ waves.  Apart from observations of the evolution of
cluster abundance, measurements of the Lyman $\alpha$ forest at high
redshift could put constraints on possible features in the power
spectrum, too.

\end{abstract}


\section{Introduction}

The distribution of matter on large scales in the Universe is supposed
to have evolved by gravitational interactions from seeds originating
in quantum fluctuations that were stretched to cosmological dimensions
by inflation. Ordinarily, this initial spectrum of fluctuations is
assumed to be a featureless power law. This requires the least number
of parameters and is produced in the simplest models of inflation.
However, different observations, such as cluster redshift surveys
(Einasto~\ea 1997) and galaxy surveys (Hamilton \& Tegmark 2000,
Gaztanaga \& Baugh 1998, Broadhurst~\ea 1990) reveal possible traces
of features in the matter power spectrum. In addition, a lower than
expected second peak in the CMB spectrum as measured by BOOMERANG
(Lange~\ea 2001) appears to be incompatible with the simplest models
of cold dark matter and a scale invariant primordial spectrum.

While the results, particularly from the surveys are still not beyond
statistical doubt, a genuine feature(s) in the primordial spectrum
cannot be ruled out.  Several mechanisms have been proposed that
could generate features in the primordial spectrum during the epoch of
inflation. They commonly involve an extension of the simplest one
field inflation model, e.g. by coupling the inflaton to a massive
particle (Chung~\ea 1999) or considering two-field inflation
(Lesgourgues, Polarski \& Starobinsky 1998). There are also more
exotic ideas (see e.g. Martin, Riazuelo \& Sakellariadou 1999).
Another idea proposed is to Taylor expand the primordial power
spectrum to include higher order terms that account for a running
spectral index (Lidsey~\ea 1997, Hannestad, Hansen \& Villante
2000). The latter is able to introduce a very broad negative or
positive bend into the power spectrum.

While possibly providing a motivation for observed features on a
specific scale (i.e. scales of about 100 Mpc), the inflationary
mechanisms proposed appear in principle capable of producing features
on other scales as well.
Here we examine the effect of primordial features on smaller scales
\mbox{$k\sim 0.4-0.8 \ h{\rm Mpc}^{-1}$},
where due to (the onset of) non-linear evolution and the problem of
biasing, a connection with the primordial spectrum is much harder to
establish. We have previously examined the effects of such bumpy power
spectra on the cosmic microwave background (Griffiths, Silk and
Zaroubi 2001).

In Section~\ref{SECpowerspectra} we will present the modifications
applied to an otherwise unperturbed SCDM power spectrum. These spectra
were then used as  input to cosmological $N$-body simulations as
described in Section~\ref{SECnbody}. Section~\ref{SECanalysis} deals
with the complete analysis of these simulations with respect to the
evolution of the power spectrum, velocity statistics, the masses of
galaxy clusters, the halo-halo correlation function, and density
profiles for a selection of halos. In Section~\ref{SECanalytical} we
try to link our numerical results to analytical prediction mainly
based on the Press-Schechter theory (Press~\& Schechter 1974). We
close with a discussion of our main results in
Section~\ref{SECconclusions}.


   \begin{figure}
	\centerline{\resizebox{\hsize}{!}{\includegraphics{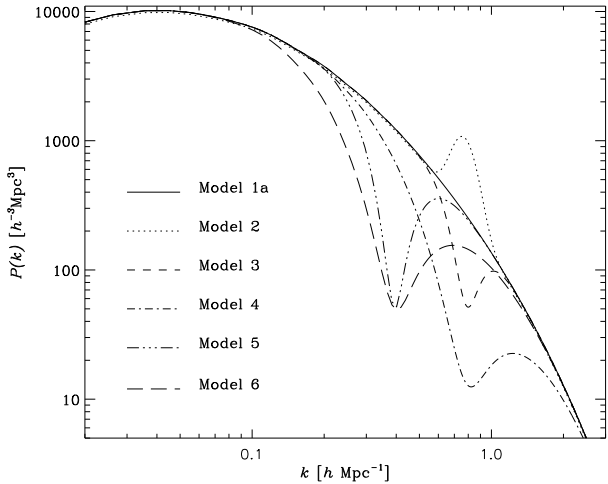}}}	
      \caption{CDM input spectra at redshift $z=0$.}
      \label{powerinput}
    \end{figure}

   \begin{figure}
      \centerline{\resizebox{\hsize}{!}{\includegraphics{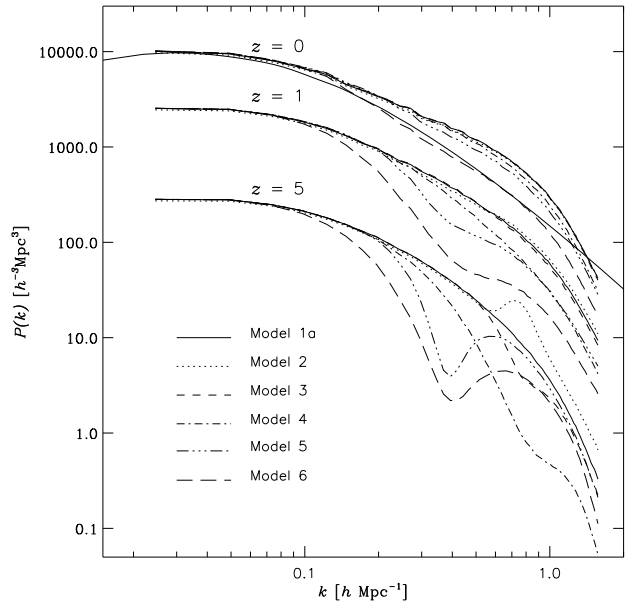}}}
      \caption{Evolution of CDM spectra.}
      \label{power}
    \end{figure}

\section{The Power Spectra} \label{SECpowerspectra}
We focus our attention on an SCDM model with parameters as given in
Table~\ref{ModelSpectra}. 
Although in this paper we are looking at properties of clusters and
particularly their abundance evolution, we decided to use only 
COBE-normalised spectra with spectral index $n = 1$ for modifications. This
is because the features we add to the spectra are on scales that at
the present time are subject to non-linear evolution and thus are
expected to lead to a deviation from the $\sigma_8 -
\Omega$ relation (Eke~\ea 1996) that would otherwise be used to {\it
cluster}-normalise the spectra.  We also focused our attention mostly
on negative amplitude features -- or dips -- in the spectrum of SCDM
models. In addition to the significance of features for structure
evolution on relatively small scales, dips will reduce the overall
power in the spectrum, pushing the normalisation and possibly other
properties towards what is expected in OCDM models.

The corresponding SCDM power spectrum was calculated using the
publicly available CMBFAST code (Seljak \& Zaldarriaga 1996) and hence
is COBE-normalised. For Model~1b we lowered the amplitude to reach a
normalisation that agrees with the cluster abundance as described in
Eke~\ea (1996).  All other models are based on Model~1a with no
further modification to the amplitude than introduced by the
artificially added feature. The modified power spectra follow
the equation:
\begin{equation}
	P_{\rm mod}(k) = P(k) \cdot (1 \pm A \exp[-0.4 (\frac{\log{k} -
	\log{k_0}}{\sigma_{\rm mod}})^2])
\end{equation}
\noindent
where P(k) is the unmodified spectrum.

These features are completely Gaussian in log-space and can be
described by their width $\sigma_{\rm mod}$, height $A$, and the
location $k_0$ of the bump (+) and dip (--), respectively.  Their
parameters are also summarised in Table~\ref{ModelSpectra}. Each
modification (characterised by the Model no. 2--6) was applied to the
SCDM spectrum corresponding to Model~1a keeping the COBE normalisation
fixed. A visual impression of the resulting power spectra can be found
in Fig.~\ref{powerinput} where all spectra are plotted (linearly
extrapolated to redshift $z=0.0$). These spectra are now used as an
input to our initial conditions generator for the cosmological \nbody\
simulations.

Again, we decided to use a COBE-normalised SCDM power spectrum as
reference model because our intention was to investigate the influence
of artificially introduced features on scales corresponding to galaxy
clusters (i.e. 8\hMpc). For this reason we chose the COBE
normalisation even though this SCDM model seems to be rather
unattractive or even ruled out nowadays. However, by taking away power
on cluster scales (as done for the majority of our 'feature' models)
the value of $\sigma_8$ drops as we do not apply any re-normalisation
of the power spectra.  The purpose was not to find a new standard
model which fits the observational data better but rather to analyse
the influence of such features on an interesting range of cluster
quantities; and this is more easily followed in a more distinctive
structure formation scenario.

\begin{table}
\caption{Specifications of the artificial modifications to the
         underlying Gaussian power spectra. The added features are of
         log-normal form, $\Lambda[e^{\mu},\sigma^2]$. All modifications
         were superimposed onto an otherwise unmodified, COBE normalised
         SCDM model (Model 1a). The corresponding mass scale $M = \rho_{\rm crit}
          \Omega \cdot \frac{4\pi}{3}(2\pi/k_0)^3$ is given in the last column.}
\label{ModelSpectra}
\begin{tabular}{llllcl}\hline
 label & $2\pi / k_0$ & $A$ & $\sigma_{\rm mod}$ & sign & mass scale \\ \hline \hline
 Model 1a  & \multicolumn{5}{c}{SCDM, $\Omega_0$=1, $h$=0.5, $\sigma_8$=1.18}\\ 
 Model 1b  & \multicolumn{5}{c}{SCDM, $\Omega_0$=1, $h$=0.5, $\sigma_8$=0.52}\\ 
 Model 2  & 8 Mpc/$h$   & 3.00 	    & 0.10   & +    & $6\cdot 10^{14}{\rm M_{\odot}}/h$\\
 Model 3  & 8 Mpc/$h$   & 0.80      & 0.12   & --   & $6\cdot 10^{14}{\rm M_{\odot}}/h$ \\
 Model 4  & 8 Mpc/$h$   & 0.95      & 0.48   & --   & $6\cdot 10^{14}{\rm M_{\odot}}/h$ \\
 Model 5  & 16 Mpc/$h$  & 0.96      & 0.24   & --   & $5\cdot 10^{15}{\rm M_{\odot}}/h$ \\
 Model 6  & 16 Mpc/$h$  & 0.96      & 0.48   & --   & $5\cdot 10^{15}{\rm M_{\odot}}/h$ \\
 OCDM     &  \multicolumn{5}{c}{$\Omega_0$=0.5, $h=0.7$, $\sigma_8$=0.96}\\ 
\end{tabular}
\end{table}

\section{The $N$-body Simulations} \label{SECnbody}
The simulations were carried out using Couchman's \ap3m\ code
(Couchman 1991). All simulations were performed with $128^3$ particles
in a box of side length 256\hMpc; the (comoving) force resolution was
fixed at 100\hkpc\ for all runs. We evolved the particle distribution
from redshift $z = 30.0$ until $z = 0.0$ in 5000 steps.
The box size was chosen such that the scales on which the Gaussian
features had been added lie comfortably within the range covered by
the simulations, e.g. our simulations cover the k-range from $k_{\rm
min} = 0.0245$ (limit set by box size) to $k_{\rm max} = 1.571$ (limit
set by particle number for representing the waves initially present),
and the modifications lie clearly within that range
(cf. Fig.~\ref{powerinput}). The particle-particle summation part of
the \ap3m\ code guarantees that we  properly
follow the evolution of all initially
present waves especially for models~2--4 (cf. Couchman 1991).

For identifying particle groups within our numerical simulations we
used the standard friends-of-friends algorithm (Davis \ea 1985) with
the linking-lengths $ll = 0.2$ for the SCDM, and $ll = 0.17$ for the
OCDM model, respectively (cf. Knebe~\&~M\"uller 1999).

\section{Analysis} \label{SECanalysis}

   \begin{figure}
      \centerline{\resizebox{\hsize}{!}{\includegraphics{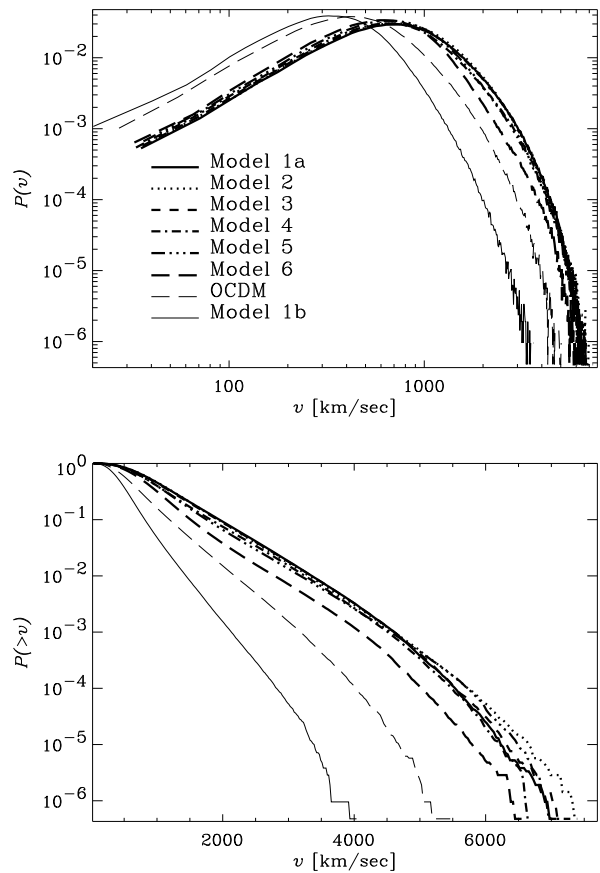}}}
      \caption{Velocity distribution for all dark matter particles}
      \label{Nv}
    \end{figure}

   \begin{figure}
      \centerline{\resizebox{\hsize}{!}{\includegraphics{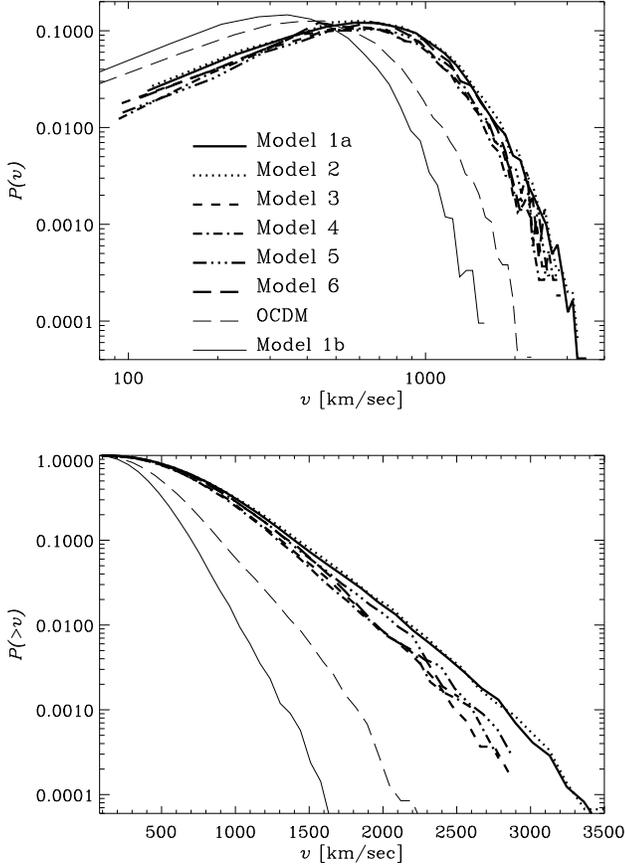}}}
      \caption{Velocity distribution for particle groups identified using
               a standard friends-of-friends algorithm with linking length 0.2.
               Only particles groups heavier than $2 \cdot 10^{13}$\hMsun\
               are taken into account.}
      \label{clNv}
    \end{figure}

\subsection{Power Spectrum Evolution}
In Fig.~\ref{power} we plot the evolution of the dark matter power
spectrum for models~1 through 6 (Model~1b and the OCDM model are left
out for clarity as there is nothing unexpected to observe).
We see that the spectrum appears to be converging towards the
unperturbed spectrum at late times.  The feature nearly vanishes
completely, leaving no further imprint in the power spectrum. The thin
solid line crossing the whole plot corresponds to the unmodified,
 COBE-normalised SCDM spectrum linearly extrapolated to $z=0.0$. Comparing
it with the spectra derived from the numerical simulations, we notice a
significant boost in power on small scales due to non-linear
evolution.  However, Model~6 with the very prominent dip at 16
\hMpc\ nearly matches the linearly extrapolated SCDM power
spectrum: at late times the non-linear boost of power almost exactly
compensates the lacking power at the location of the dip.

We may conclude that even when starting with a prominent feature on
small scales, there might only be little if any evidence for it left
in the present universe when looking at the dark matter power spectrum
P(k).  In any case, the {\em evolution} of the models {\em has} to be
different, and one needs to think of other ways to investigate the
influence of such features in the present day universe or to detect
them in the evolution of P(k) and related quantities.  As far as the
spectrum itself is concerned current galaxy clustering surveys offer
little prospect of detecting these features in the evolution of
P(k). Observations of the Ly$\alpha$ forest, however, appear to be a
promising tool for constraining P(k) at high redshift (see
e.g. Croft~\ea 2000, Weinberg~\ea 1998), particularly on the scales
under investigation here.

\subsection{Velocity Statistics}
Next we consider the distribution of velocities for a)~the dark matter
particles and b)~friends-of-friends particle groups. The peculiar
velocity field results from the gravitational acceleration that
develops from initial density fluctuations in the early
universe. Clusters of galaxies can therefore be used as tracers of the
large-scale peculiar velocity field (Bahcall, Cen \& Gramann 1994).

In Fig.~\ref{Nv} we show the probability distribution $P(v)$ of dark
matter particles with peculiar velocities in the range $v\pm dv$ along
with the integrated distribution $P(>v)$. The same is plotted for
friends-of-friends groups identified in all runs in
Fig.~\ref{clNv}. All curves are normalised by the total number of
particles and total number of particle groups in the respective model.
Here we also show the data for SCDM Model~1b as well as the fiducial
OCDM model. Even though there are differences of a factor of two at
the high velocity end of the distribution, it might be difficult to
observationally distinguish  these differences between modified and
 unmodified models. 

The deviations between the cluster-normalised SCDM model and the
COBE-normalised one are bigger than any differences between our
'feature' models. This deviation in velocities tending to lower values
in the cluster-normalised model can be explained by the lower value
for $\sigma_8$: the amplitudes of the initial density fluctuations are
smaller and hence it takes longer to accelerate particles (and
clusters) to high velocities; in the course of a simulation the
velocity distribution function is always similar to a Maxwellian
distribution whose peak gradually moves from low to high
velocities. Apart from this we get increasingly more high velocity
particles leading to a bigger 'tail' in the distribution. However, the
most prominent difference is in the high velocity tail of the
distribution for clusters. Only here it is possible to discriminate
between models as our dip models show a significant drop in the
integrated probability. We do not observe any galaxy clusters with
peculiar velocities higher than 3000 \kms\ whereas there is a
distinctive number of these objects in the unmodified Model~1a (and
the bump Model~2). A detailed check showed that this tail is mainly
due to objects with masses $M \lesssim 10^{14}$\hMsun. When only taking
into account particle groups with heavier than $10^{14}$\hMsun all
curves for the 'feature' models fall on top of each other.  However,
the location and the width of the feature does not seem to have any
significant influence.

\subsection{The Masses of Galaxy Clusters}
The most basic property of a galaxy cluster is its mass
$M$. Nevertheless, this quantity can provide a lot of information
especially when using the (cumulative) distribution of objects with a
certain mass $M$; and the evolution of the abundance of massive clusters
 within
a given mass range is indeed one of the corner stones of the currently
favoured \LCDM\ model (Bahcall \ea 1999).

\subsubsection{Mass Function and Press-Schechter Prediction}
\begin{figure}
      \centerline{\resizebox{\hsize}{!}{\includegraphics{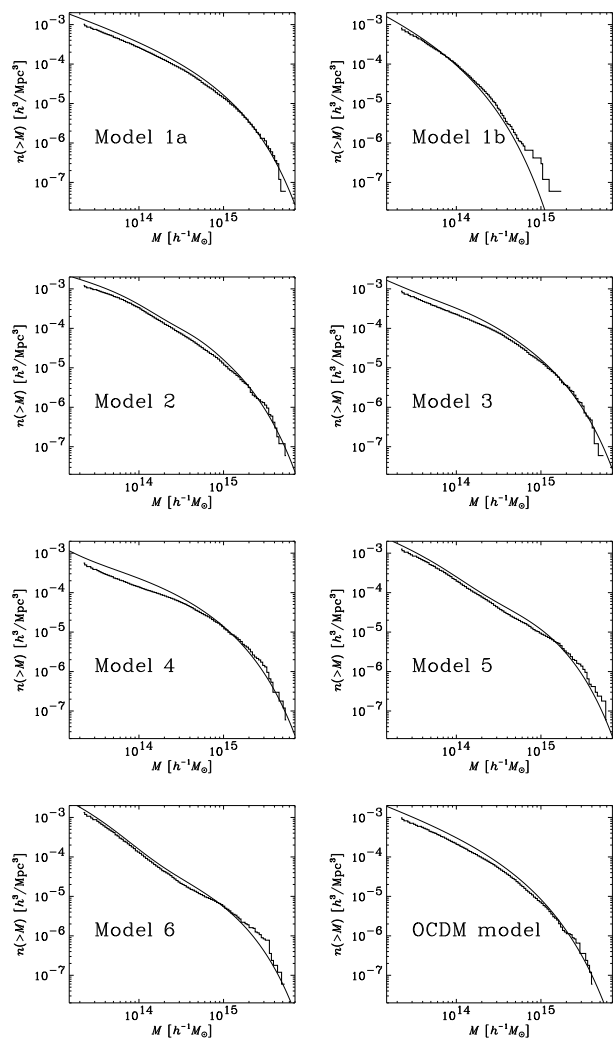}}}
      \caption{CDM mass function for redshifts $z=0.0$ from
               simulations (histograms) and Press-Schechter prediction (thin lines).}
      \label{mass}
    \end{figure}

In Fig.~\ref{mass} we show the cumulative mass function of particle
groups for all our models at a redshift of $z=0.0$.  Even though we
could not find well-pronounced imprints of the modifications in the
power spectra at redshift $z=0.0$ (cf. Fig.~\ref{power}), we clearly
see differences in the amplitude and slope of the cumulative mass
function $n(>M)$. These deviations are mainly at the low mass end of
the resolvable mass range, where a positive feature leads to an
excessive number of objects and a negative feature to a lack of
groups. This is in general agreement with an excess/lack of objects on
mass scales corresponding to where the features are located.  However,
we always end up with the same number of galaxy clusters for masses $M
> 10^{15}$\hMsun (besides for Model~1b which shows too few massive
groups due to the low normalisation).
Unfortunately we are not able to resolve particle groups lighter than
about $2\cdot 10^{13}$\hMsun\ and moreover, the simulation volume
is still too small to get a statistically significant number of
objects heavier than about $3\cdot 10^{15}$\hMsun.

  \begin{figure}
      \centerline{\resizebox{\hsize}{!}{\includegraphics{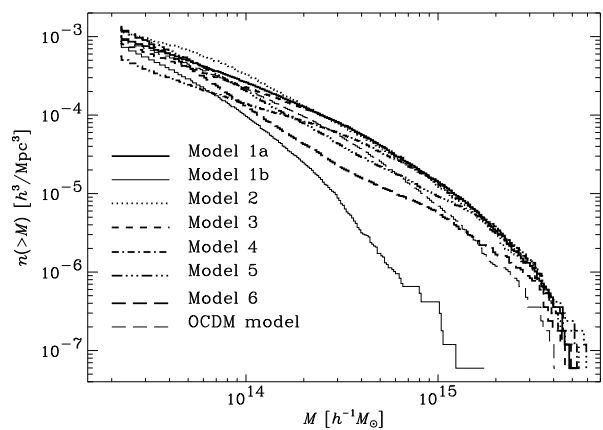}}}
      \caption{CDM mass functions as presented in Fig.~\ref{mass}.}
      \label{mass2}
    \end{figure}

To check the validity of our results and get an impression of how they
might generalise for larger/smaller mass objects we performed
Press-Schechter (PS) calculations (Press~\& Schechter 1974) of the
abundance of gravitationally bound objects.
The (differential) number of objects for a given mass $M$ can be
calculated using their formula:
\begin{equation} \label{PS}
 \displaystyle \frac{dn}{dM} dM = 
 \sqrt{\frac{2}{\pi}} \frac{\overline{\rho}}{M} \frac{\delta_c}{\sigma_M}
                    \left|\frac{d \ln \sigma_M}{d \ln M}\right|
                    \exp{\left(-\frac{\delta_c^2}{2 \sigma_M^2}\right)}
                    \frac{dM}{M}
\end{equation}
\noindent
with the variance $\sigma_M$ defined as follows:
\begin{equation} \label{sigmaM}
 \sigma_M = \int P(k) W^2(k R) k^2 dk \ ,
\end{equation}
\noindent
where $W(kR)$ is the window function (top-hat in our case) for
filtering fluctuations in the power spectrum on scales characterised
by $R$ and hence mass $M = 4\pi R^3/3$. Using our model power spectra
from Table~\ref{ModelSpectra} (plotted in Fig.~\ref{powerinput})
together with Eq.~(\ref{sigmaM}) and Eq.~(\ref{PS}) we are able to
compare our numerically achieved mass functions with the
PS~prediction. The results are shown in Fig.~\ref{mass} as thin solid
lines. 

We observe a similar phenomenon as already seen in other comparisons
of PS-predicted and $N$-body mass functions (Efstathiou~\ea 1988,
White, Efstathiou~\& Frenk 1993, Gross~\ea 1998, Governato~\ea 1999,
Jenkins~\ea 2001): the PS theory tends to show too many low mass
objects (about a factor of 1.2) and too few high mass objects (again a
factor of about 1.2).  Apart from that the simulations agree fairly
well with the Press-Schechter prediction, even in the cases where we
modified the power spectra.

Anyway, to allow for better comparison between the individual models
and the effect of the features on $n(>M)$ we also plot all mass
functions derived from the numerical simulations (as already presented
in Fig.~\ref{mass}) in one single Figure~\ref{mass2}.

\subsubsection{Evolution of the Cluster Abundance} \label{evolution}
   \begin{figure}
      \centerline{\resizebox{\hsize}{!}{\includegraphics{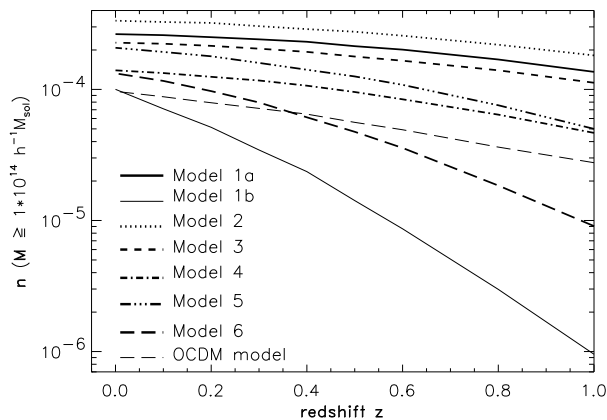}}}
      \caption{Evolution of SCDM cluster abundance.}
      \label{abundance}
    \end{figure}

   \begin{figure}
      \centerline{\resizebox{\hsize}{!}{\includegraphics{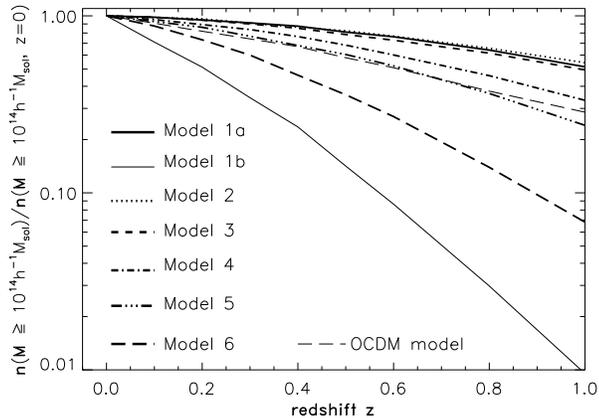}}}
      \caption{Evolution of SCDM cluster abundance normalised to unity at
               redshift $z=0$.}
      \label{abundance2}
    \end{figure}

The evolution of the cluster abundance has proven to be (potentially)
one of the key constraints on the density parameter $\Omega_0$
(e.g. Eke \ea 1996, Bahcall \ea 1997, Eke \ea 1998, Bahcall
\ea 1999). This leads immediately to the question of how our
modifications to the dark matter power spectrum influence this
important issue. We have already seen that while
we might not find hints in
the observational power spectrum for the features under investigation,
 there are differences in the mass functions $n(>M)$.  In
Fig.~\ref{abundance} we show the evolution of galaxy clusters with
mass~$M$ greater than $10^{14}$\hMsun.

The first thing that catches the eye is the difference between the
COBE and cluster-normalised SCDM model when  compared to the OCDM
model. Only the cluster-normalised SCDM model shows a very steep
evolution of the cluster abundance whereas the evolution in the
 COBE-normalised one agrees more or less with the OCDM model (cf. Bahcall,
Fan \& Cen 1997, Fig.~2 in their paper). It is now interesting to
check whether our features affect only the normalisation of these
evolutionary curves or also the slope.  As we can see, the
variations in the amplitude of the dip (or bump) mainly shift the
cluster evolution up and down, and the location has an influence
primarily on the slope of the curve. 

This can be understood in the
following way. Features on the 8\hMpc\ scale are 'dynamically' important, in that
this scale has evolved from the linear regime ($z \sim 1$) into
quasi-linear or non-linear regime by now. Since the relative abundance
of objects increases dramatically in the non-linear regime, we
consequently expect still more (less) objects at late times if power
had been added (subtracted) on the corresponding scale, i.e. features
on this scale are expected to affect the shape of the abundance
evolution of corresponding mass objects. Conversely, features on
scales that up to present have remained in the linear regime only add
(subtract) power on dynamically unimportant scales, thus only
affecting the normalisation.
\footnote{Though, if the difference in power is very large (c.f. abundance
evolution for $\sigma_8 = 1.18$ and $\sigma_8 = 0.52$) there is a
significant change in the slope of the abundance evolution, too.}

Moving the dip from the non-linear scale of 8\hMpc\ to the semi-linear
16\hMpc\ mainly results in an overall lack of power and therefore we
expect the cluster evolution to be closer to the cluster-normalised
($\sigma_8=0.52$) Model~1b. However, the \textit{evolution} for
Model~4 agrees fairly well with the OCDM Model. This can be seen even
better in Fig.~\ref{abundance2} where we normalised the cluster
abundance at redshift $z=0$ to unity for all models to allow for
better comparison of the slopes of the curves.
 \begin{figure}
      \centerline{\resizebox{\hsize}{!}{\includegraphics{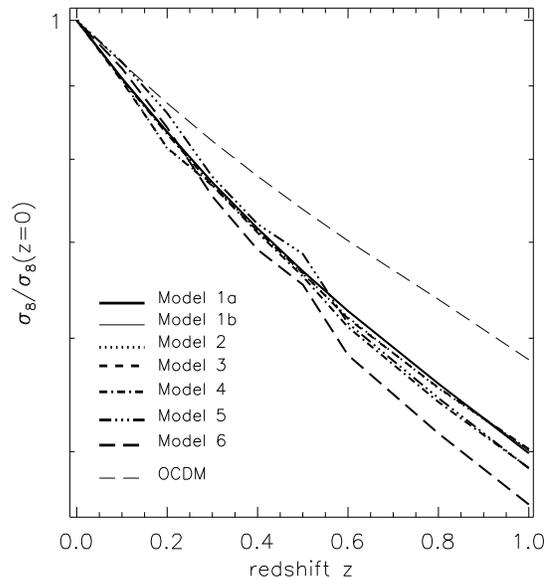}}}
      \caption{Evolution of $\sigma_8$ in all models. The evolution in
               all models is normalised to be unity at redshift $z=0$.}
      \label{sigmaR2}
    \end{figure}

A similar effect was already observed by Barriga~\ea (2000) where the
influence of a step-like feature in the primordial power spectrum was
studied in the context of phase transitions during inflation.
However, their features are on larger scales and thus would leave
traces in current epoch measurements of the CMB and large scale
structure.

\subsection{$\sigma_8$ Evolution}
As the number of objects formed depends on the normalisation of the
input power spectrum measured via $\sigma_8$ (cf. Eq.~\ref{sigmaM}),
it is interesting to check the evolution of this quantity with
redshift, too.  Fig.~\ref{sigmaR2} shows this evolution for all
models. Again, all curves are normalised to the value at redshift
$z=0$ to allow for better comparison of the slopes.
We observe no overlap of any feature model with the OCDM
model: regarding the evolution of the dark matter fluctuations
measured in 8\hMpc\ spheres (rather than the cluster evolution) the
OCDM model still differs significantly from all other
models. Moreover, it is difficult again to distinguish between our
feature models as well as the differently normalised SCDM models
itself. Only Model~6 shows a steeper evolution from redshift $z \sim
1$ to $z \sim 0.5$. 
Hence $\sigma_8$-evolution  does not provide a discriminant
of features in the power spectrum.
However, Robinson, Gawiser and Silk (2000) have shown that
$\sigma_8$ and cluster abundance combine to probe non-gaussianity.

\subsection{Halo-Halo Correlation}

\begin{figure}
      \centerline{\resizebox{\hsize}{!}{\includegraphics{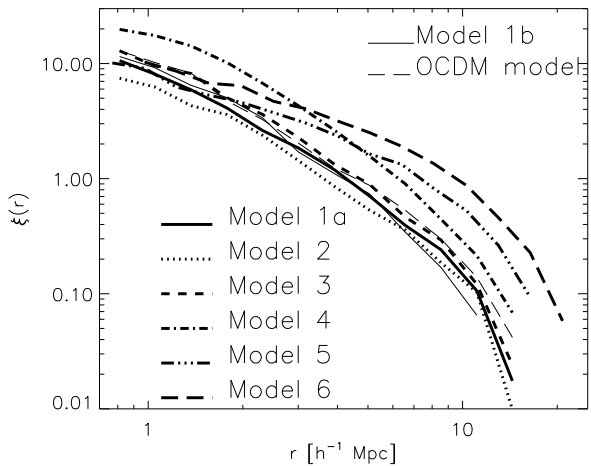}}}
      \caption{Halo-Halo correlation function for particle groups.
               The number density of objects was fixed in all 
               models to $n = 5 \cdot 10^{-4} h^3 {\rm Mpc}^{-3}$.}  
      \label{halohaloXi}
      \end{figure}

In Fig.~\ref{halohaloXi} we present the halo-halo correlation function
for our models. We fixed the number density to $n = 5 \cdot 10^{-4}
h^3 {\rm Mpc}^{-3}$ which actually means to only use the $N = n \cdot
V$ most massive halos with $V=256$\hMpc\ being our simulation volume.

We now observe more obvious differences between these models,
which can be fully ascribed to our artificial modifications of the
initial power spectrum.  The amplitude of the cluster correlation
function is actually insensitive to the amplitude of fluctuations in
the density field (Croft \& Efstathiou 1994), which is clearly
reflected when comparing Model~1a and 1b.  Moreover, there are also
only moderate changes in the cluster correlation function when varying
the cosmological parameters from our SCDM Model~1a/b to the OCDM model
(cf. Martel \& Matzner 2000). But as we move on to our feature models
not only the amplitude of the correlation function increases (for the
dip models), but also the slope changes (i.e. Model~6). This suggests
a strongly (and non-linearly) biased formation of galaxy clusters in
those models compared to a 'normal' SCDM model.

As there was no obvious biasing observed in the dark matter power
spectrum (which is nothing more than the Fourier transform of the dark
matter correlation function), the effect seen in Fig.~\ref{halohaloXi}
is entirely due to a different cluster formation scenario in our dip
models. This is caused by the fact that by adding (subtracting) power
only on a certain scale the relative strength of different waves
becomes more and more important. If we put less power into high-$k$
waves (i.e. Model~3 and~4) but leave the largest waves unmodified we
eventually bias the formation process of galaxy clusters with respect
to an unmodified SCDM model as can be observed in
Fig.~\ref{halohaloXi}. Massive clusters indeed are highly biased, and
the extent to which one may have difficulty in accounting for this in
standard low density models may constitute the strongest signature of
a possible feature at redshift $z=0$ in a SCDM model.

\subsection{Density Profiles}

   \begin{figure}
      \centerline{\resizebox{\hsize}{!}{\includegraphics{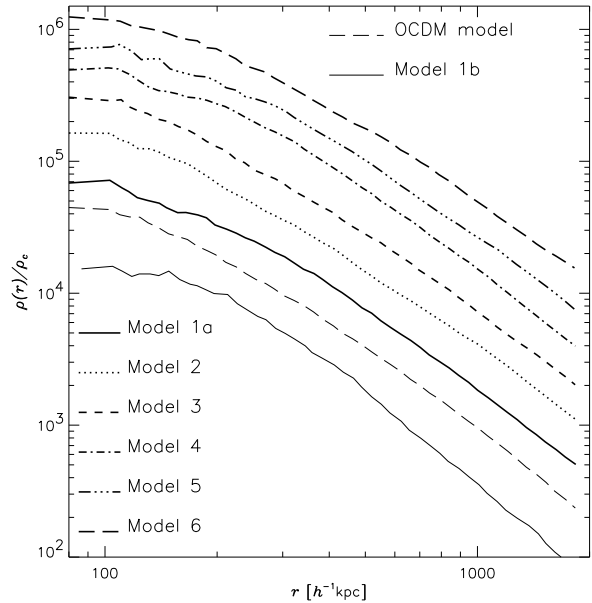}}}
      \caption{Density profile for one halo in all models.}
      \label{profile}
    \end{figure}

We have already seen that we introduced a strong bias in the cluster
formation process by altering the ratio of high- to low-$k$ waves
amplitudes. This immediately raises the question of whether this will
subsequently lead to deviations in the shapes of the clusters
themselves.  We have therefore calculated the density profiles for
massive FOF groups ($M > 10^{15}$\hMsun) and show the results for a
representative high-mass halo $M \sim 3.5\cdot 10^{15}$\hMsun\
($\sim$~1700 particles) in Fig.~\ref{profile}.  The curves for
models~2 through~6 have been shifted upwards by successive factors
of~2, whereas the OCDM profile was lowered by a factor of 2.

We can see that modifying the power spectrum does not affect the slope
and amplitude of the density profiles. Only the SCDM model~1b deviates
from the other curves as the corresponding halo consists of only about
500 particles. Due to the resolution limits of our simulations, we are
not able to quantify the substructure content of these galaxy clusters
in more detail as the most massive objects contain 'only' 2000-2500
particles. But a visual comparison of the particle distributions in
those clusters provided no obvious differences.

\section{Analytical Estimates} \label{SECanalytical}

We have seen above that SCDM models with features at small scales
may display an evolution of their cluster abundance that is similar to
that in the OCDM model to within some constant offset
in amplitude.  The question remains of how far we can push this agreement
and to what extent this also holds for cluster masses larger than the ones 
we were able to investigate numerically.  
To this end we performed Press-Schechter (Press \&
Schechter 1974) calculations to investigate a larger set of
models. However, since we are considering the evolution of cluster
abundances, this also requires a relation for the redshift dependence
of the critical overdensity $\delta_c$, which is left unspecified in
the original PS approach.  We determined approximate relations for the
unmodified SCDM and OCDM models by matching their cluster abundance
evolution to that of the corresponding N-body simulations:
\begin{equation}\label{deltaeq}
\delta_c^{SCDM}(a) = 1.2 + 0.6a \hspace{1pt} ,  \\   
\delta_c^{OCDM}(a) = 1.45 + 0.3a 
\end{equation}
\noindent
where $a = 1/(1+z)$ is the cosmic scale factor at redshift z.
 \begin{figure}
      \centerline{\resizebox{\hsize}{!}{\includegraphics{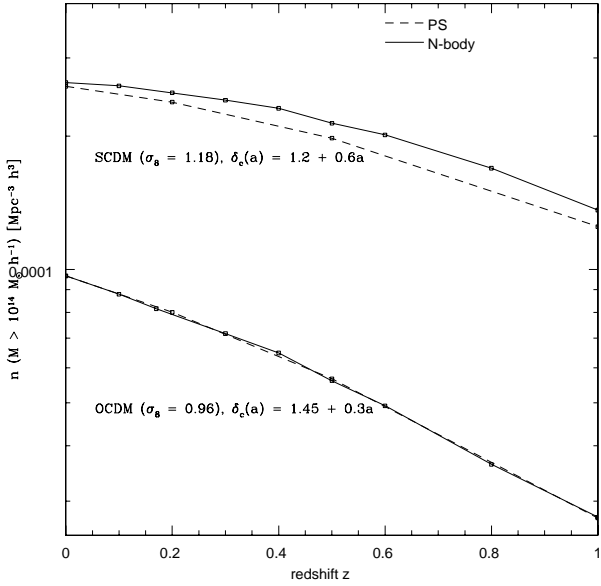}}}
      \caption{Comparison of evolution of clusters of mass 
               $M > 10^{14} M_{\odot} h^{-1}$ in simulations and PS
      using the relations for $\delta_c$ from the simulations.}
      \label{dccomp}
    \end{figure}

   \begin{figure}
      \centerline{\resizebox{\hsize}{!}{\includegraphics{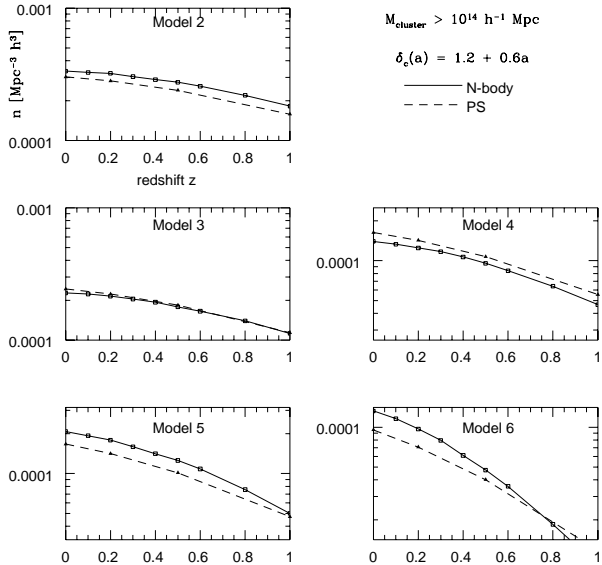}}}
      \caption{Comparison of abundance evolution for clusters with mass  
               $M > 10^{14} M_{\odot} h^{-1}$ in simulations and
      PS. All PS calculations use $\delta_c(a) = 1.2 + 0.6a$ (eq (\ref{deltaeq})).}
      \label{dccompall}
    \end{figure}
For the OCDM model the $\delta_c$ relation almost exactly
      reproduces the abundance evolution. The corresponding relation
      for the SCDM model was rather chosen to fit both unmodified SCDM and
      featured SCDM models reasonably well, matching the N-body
      evolution to within 10 percent for the unmodified SCDM model,
      typically to within 20 percent for models 2 to 5 and 30 percent for
      model 6. This is shown in figures ~\ref{dccomp} and
      ~\ref{dccompall}. An accurate fit to the unmodified
      SCDM model only would have required $\delta_c(a) = 1.25 + 0.5a$.


However, the agreement between N-body and PS data should be viewed
with care, particularly when large sharp features are included:
At the location of the features, we essentially introduce a more
rapidly changing spectral index.  This might contribute to a larger
discrepancy between N-body and PS data.  To our knowledge agreement
between PS and numerical simulations has so far only been established for
power spectra with constant (power law) or slowly varying (e.g. SCDM)
spectral indices - it is not obvious why this should also hold when a
sharply varying index is being introduced, particularly on scales that
only now turn non-linear.
In the following, however, we will assume that this is not a problem
and use the $\delta_c^{SCDM}$ relation from equation (\ref{deltaeq}) as a best
estimate for PS predictions of the abundance evolution for other
small scale features in the SCDM model. We also assume that the $\delta_c$ relation for
OCDM describes the abundance evolution in higher mass cuts as well
as it does for $M > 10^{14}$ \hMsun in our simulations. 

As seen above, one might expect that adding (subtracting) power only
at specific scales leads to a larger (smaller) abundance of objects
with mass corresponding to these scales. This is not obvious for
scales that are already in the non-linear regime, as we have a
coupling across scales, and also objects not only get newly created
but now have also been partly incorporated into larger objects,
i.e. the effect of a feature on a specific scale spreads out, not only
in the evolution of the power spectrum as we have seen above, but also
in the abundance of objects of corresponding mass.  However, for broad
features we may assume that the latter is less important, since a
whole range of scales and corresponding masses are affected.

To check this, we placed a broad Gaussian dip in the SCDM model ($A =
-0.55$, $\sigma_{\rm mod} = 2.5$) at 10\hMpc, which is expected to
affect a similarly broad range of masses centred on $M \sim 10^{14} -
10^{15} M_{\odot} h^{-1}$. 
A broad feature of this kind is naturally generated by e.g. including
a running spectral index term when expanding the primordial
perturbation spectrum (Hannestad, Hansen \& Villante 2000).
  \begin{figure}
      \centerline{\resizebox{\hsize}{!}{\includegraphics{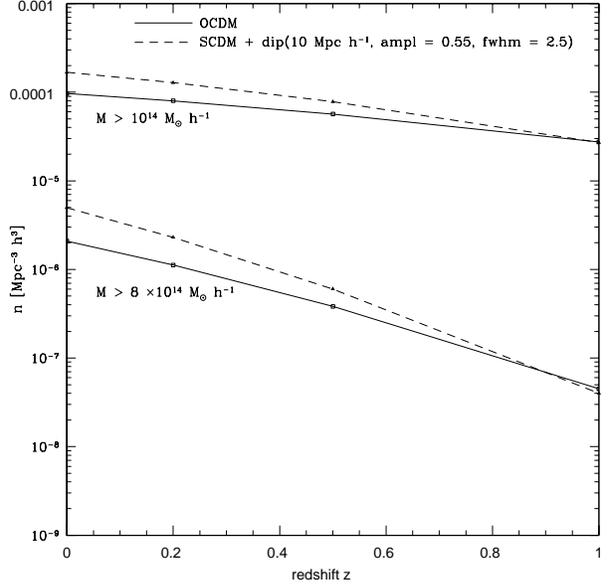}}}
      \caption{PS calculations of abundance evolution for OCDM and
      SCDM with a broad dip at 10 Mpc $h^{-1}$. For the mass
      cuts the two agree to within less than a factor of 2 and 3
respectively across the whole redshift range.}  \label{OvsSdip} \end{figure}

Comparing Fig.~\ref{dccomp} with Fig.~\ref{OvsSdip} confirms that the
abundance of masses in these ranges ($M > 10^{14}$, $M > 8 \cdot
10^{14}$\hMsun, the mass range covered by e.g. Bahcall \& Fan 1998) is
significantly suppressed. What is
more, the abundance evolution for the mass cuts agrees in logarithmic slope {\it
and} magnitude with that in the OCDM model to within a factor of less 
than two and three across all redshifts up to z = 1 for the lower
and upper mass  cuts respectively. 



Given that observations of the most distant massive clusters currently
only constrain the abundance to within one or two orders of magnitude
(see e.g. Bahcall \& Fan 1998), SCDM with broad features towards small
scales, mimicking (in shape and absolute magnitude) the
abundance evolution expected in OCDM models, can therefore not be
ruled out. A similar result is obtained by Barriga {\it et al.}(2001)
who consider primordial step-like features in the context of phase
transitions during inflation, however their features are on larger
scales in the linear regime and thus would also leave traces in current
epoch measurements of the CMB and large scale structure.

\section{Conclusions} \label{SECconclusions}

In this paper we presented a series of simulations all based on the
COBE normalised SCDM model but with Gaussian features added to an
otherwise unperturbed power spectrum (called 'feature' models).  Such
bumps (or dips) might naturally arise from non-standard inflationary
theories and our main purpose was to investigate their influence on
the large-scale structure of the Universe as measured via galaxy
clusters. We analysed the evolution of the dark matter power spectrum,
the large-scale velocity field represented by the velocity
distribution of galaxy clusters, the evolution of the cluster
abundance, the halo-halo correlation function, and density profiles of
clusters. We furthermore compared our numerical results to analytical
predictions based on the Press-Schechter formalism (Press~\& Schechter
1974).

When comparing the modified SCDM models to a fiducial OCDM model, we
can see that when choosing an appropriate scale for the added feature
the histories  of cluster abundance evolution might be
indistinguishable. However, there is still
a discrepancy in the overall normalisation left at the
present epoch,
 with only the slopes of the cluster abundance evolution coinciding.
But as  observations of the most distant massive
clusters only constrain the current abundance to within one or two orders of
magnitude (cf. Bahcall~\& Fan 1998), an SCDM model including such a
broad feature cannot be ruled out.

Finally we remark that whatever the origin or nature is
of these bumps and dips,
their effect on the evolutionary history of the SCDM model is much
more moderate than changes in the cosmological parameters of the model
itself, e.g. lowering the normalisation from $\sigma_8=1.18$ (COBE
normalisation) to $\sigma_8=0.52$ (cluster normalisation).  It is
therefore an observational challenge to find traces of such features,
and the best place to search for them might be the cluster-cluster
correlation function which appears to be highly biased with respect to
the 'normal' SCDM model. This is due to an unusual ratio of power for
high- and low-$k$ waves, which will be reflected in the aforementioned
'more than biased' cluster formation scenarios.

\section*{Acknowledgements}

We are grateful to the referee Mirt Gramann for helpful comments and
suggestions. We also
acknowledge the use of Hugh Couchman's \ap3m\ code. RRI gratefully
acknowledges a Graduate Studentship from Oxford University and support
from St Cross College, Oxford.


\end{document}